\documentstyle[12pt]{article}
\def\beq{\begin{equation}}
\def\eeq{\end{equation}}
\def\bea{\begin{eqnarray}}
\def\eea{\end{eqnarray}}
\def\bq{\begin{quote}}
\def\eq{\end{quote}}
\def\PL{{ \it Phys. Lett.} }
\def\PRL{{\it Phys .Rev. Lett.} }
\def\NP{{\it Nucl. Phys.} }
\def\PR{{\it Phys. Rev.} }

\def\gappeq{\mathrel{\rlap {\raise.5ex\hbox{$>$}}
{\lower.5ex\hbox{$\sim$}}}}

\def\lappeq{\mathrel{\rlap{\raise.5ex\hbox{$<$}}
{\lower.5ex\hbox{$\sim$}}}}
\begin{document}
\renewcommand{\thesection}{\Roman{section}.}
\renewcommand{\theequation}{\arabic{section}.\arabic{equation}}
\pagestyle{empty}
\begin{flushright}
CERN-TH.7227/94\\
NUB-TH-3089/94\\
CTP-TAMU-28/94\\
\end{flushright}
\vspace*{2.0cm}
\begin{center}
SUSY PARTICLES\footnote{Invited Talk at the Les Rencontres de Physique
de la Vall\'ee
d'Aoste, ``Results and Perspectives in Particle Physics", at La Thuile,
Aosta Valley,
Italy, 6--12 March 1994.}

Pran Nath\\
Theoretical Physics Division, CERN\\
1211 Geneve 23, Switzerland\\
and\\
Dept. of Physics, Northeastern University\\
Boston MA 02115, U.S.A.\footnote{Permanent address.}\\
\vspace{0.5cm}
and\\
\vspace{0.5cm}
R. Arnowitt\\
Center for Theoretical Physics, Dept. of Physics,\\
College Station TX 77843, U.S.A.\\
\vspace*{2.0cm}
Abstract
\end{center}
Analysis of the SUSY spectrum in supergravity unified models is given
under the
naturalness criterion that the universal scalar mass $(m_0)$ and the
gluino mass
$(m_{\tilde g})$ satisfy the constraint $m_0, m_{\tilde g} \lappeq
1$~TeV.  The SUSY
spectrum is analysed in four different scenarios:  (1) minimal
supergravity models
ignoring proton decay from dimension five operators (MSSM), (2) imposing
proton
stability constraint in supergravity models with SU(5) type embedding
which
allow proton decay via dimension five operators, (3) with inclusion of
dark matter
constraints in models of type (1), and (4) with inclusion of dark matter
constraint
in models of type (2).  It is found that there is a very strong upper
limit on the
light chargino mass in models of type (4), i.e., the light chargino mass
$\lappeq$~120 GeV.
\vspace*{1.0cm}
\begin{flushleft}
CERN-TH.7227/94\\
NUB-TH-3089/94\\
CTP-TAMU-28/94\\
April 1994\\
\end{flushleft}

\vfill\eject
\pagestyle{empty}
\clearpage\mbox{}\clearpage
\setcounter{page}{1}
\pagestyle{plain}

\section{Introduction}

Supergravity unification$^{1,2)}$ is currently the leading candidate
theory for
physics beyond the Standard Model.  It allows for a phenomenologically
viable
breaking of supersymmetry$^{2,3)}$, and the formalism also generates its
own breaking
of the electroweak symmetry via renormalization group effects.  The
search for
supersymmetric particles is of great interest at supercolliders, in dark
matter
experiments and elsewhere.  Here we discuss the computation of the SUSY
mass
spectrum in four specific scenarios.  These consist of supergravity
models
excluding or including proton stability constraint, and supergravity
models with
and without dark matter constraints.  The models where we impose proton
stability are
those which have an SU(5) type embedding.

The outline of the paper is as follows:  in Section~2 we discuss general
features of
SUSY models.  In Section~3 we discuss supergravity models, the parameter
space of the theory, and the radiative breaking of the electroweak
symmetry and constraints that are imposed in the analysis of the SUSY
spectrum.  In
Section~4 we give the renormalization group analysis of SUSY parameters
and analytic
formulae for the SUSY spectrum.  Section~5 is devoted to analysis and
results of the
spectrum for four different scenarios.  Conclusions are given in
Section~6.

\section{SUSY models:  particles and interactions}
\setcounter{equation}{0}
SUSY models are built using two types of massless supermultiplets;
chiral
multiplets with spin (0,1/2) is a left-handed chiral multiplet, and
vector
multiplets with spin (1/2,1) where spin 1/2 is a Majorana spinor and
spin 1 is a
vector boson.  For SUSY extension of the standard SU(3)$_{\rm C} \times$
SU(2)$_{\rm
L} \times$ U(1)$_{\rm Y}$ model, the chiral multiplets consist of the
following:
\begin{equation}
\begin{array}{lcc}
&\,\, J=0 & \,\, J=1 \\
\begin{array}{c} \mbox{ lepton  } \\ \left. \mbox{multiplet} \right.
\end{array}
&\!\!\left( \begin{array}{c}
\!\!\tilde \nu_{iL}\\
\!\!\tilde e_{iL} \end{array} \!\!\right) \begin{array}{c}\mbox{, }
\tilde e_{iR} \\
\left. \right.\end{array}& \!\!\left( \begin{array}{c}
\!\!\nu_{iL}\\\!\! e_{iL} \end{array}
\!\!\right)\begin{array}{c}\mbox{, } e_{iR}\\ \left.\right.
\end{array}\\*[20pt]
\begin{array}{c} \mbox{ quark } \\ \left. \mbox{multiplet} \right.
\end{array}
&\!\! \left( \begin{array}{c}
\!\!\tilde u_{iL}\\
\!\!\tilde d_{iL} \end{array}\!\! \right) \begin{array}{c}\mbox{, }
\tilde u_{iR}~, \tilde
d_{iR} \\ \left. \right.\end{array}
&\!\! \left( \begin{array}{c}u_{iL}\\
\!\!d_{iL} \end{array}\!\! \right)\begin{array}{c}\mbox{, }
u_{iR}~,d_{iR}\\ \left.\right.
\end{array}\\*[20pt]
\begin{array}{c} \mbox{ Higgs  } \\ \left.\mbox{multiplets} \right.
\end{array} &
\begin{array}{c} H_1= \\ \left.\right. \end{array}
\!\!\!\!\left( \begin{array}{c}
\!\!H^0_1\\
\!\!H^-_2 \end{array} \right) \begin{array}{c}\mbox{, }
H_2= \\ \left.\right. \end{array}
\!\!\!\!\left( \begin{array}{c}
\!\!H^+_2\\
\!\!H^0_2 \end{array} \right) &
\begin{array}{c} \tilde H_1= \\ \left.\right. \end{array}
\!\!\!\!\left( \begin{array}{c}
\!\!\tilde H^0_1\\
\!\!\tilde H^-_1 \end{array} \right) \begin{array}{c}\mbox{, }
\tilde H_2= \\ \left.\right. \end{array}
\!\! \!\!\left( \begin{array}{c}
\!\!\tilde H^+_2\\
\!\!\tilde H^0_2 \end{array} \right)
\end{array}
\label{21}
\eeq
The potential of this theory is given by
\beq
V = V_W + V_D + V_{SB}
\label{22}
\eeq
where $V_W = \sum |\partial W / \partial z_a|^2$ and $W$ is the
superpotential
given by
\beq
W = \mu H_1H_2 + \lambda^{(e)}_{ij} \ell_iH_1e^c_j +
\lambda^{(u)}_{ij}q_iH_2u^c +
\lambda^{(d)}_{ij}q_iH_1d^c_j
\label{23}
\eeq
In Eq.~(\ref{22}) $V_D$ is the $D$-term given by $V_D = \frac{1}{2}
g^2_AD_AD^\dagger_A$ with $D_A = z_a^+(T^A)_{ab}z_b$, and $V_{SB}$ is
the SUSY breaking
terms.  It has the general form$^{4)}$
\bea
V_{SB} &=& m^2_{ab}z_a^*z_b + \left[
A^{(e)}_{ij}\lambda^{(e)}_{ij}\ell_iH_1e^c_j
+ A^{(u)}_{ij}\lambda^{(u)}_{ij}q_iH_2u^c_j \right. \nonumber \\
&& \left. + A^{(d)}_{ij}\lambda^{(d)}_{ij}q_iH_1d^c_j + B\mu H_1H_2 +
{\rm h.c.}
\right]
\label{24}
\eea
The SUSY breaking terms in Eq.~(\ref{24}) have a large number of
arbitrary
parameters.  Thus this theory is not very predictive.  We shall next
discuss
supergravity unification where there is a sharp reduction in the number
of
arbitrary parameters, and the theory is very predictive.

\section{Supergravity Unification}
\setcounter{equation}{0}
In supergravity unification supersymmetry can be broken spontaneously
via a hidden
sector, and the effective low-energy theory has only four arbitrary
parameters.
In this theory the effective potential below the GUT scale $M_G$ is
given
by$^{2,3,5,6)}$
\beq
V_{SB} = m^2_0 z_iz^+_i + \left( A_0W^{(3)} + B_0 W^{(2)} + {\rm
h.c.}\right)
\label{31}
\eeq
In Eq.~(\ref{31}) $W^{(2)}, W^{(3)}$ are the quadratic and the cubic
parts of the
effective superpotential which in general has the expansion
\beq
W_{eff} = W^{(2)} + W^{(3)} + \frac{1}{M} W^{(4)}
\label{32}
\eeq
where $W^{(2)} = \mu_0H_1H_2, W^{(3)}$ contains terms cubic in fields
and involves
the interactions of quarks, leptons and Higgs, and $W^{(4)}$ contains
terms quartic
in fields and in general has interactions which violate baryon number.
In addition
to SUSY breaking terms, in Eq.~(\ref{31}) one also has a universal
gaugino mass
term of the form $m_{1/2} \bar \lambda^\alpha \lambda^\alpha$.  Thus the
effective
theory below the GUT scale depends on the following set of parameters:
\beq
m_0, m_{1/2}, A_0, B_0~;~~\mu_0~;~~\alpha_G,~M_G
\label{33}
\eeq
where $M_G$ is the GUT mass and $\alpha_G$ is the GUT gauge coupling
constant.

Below the GUT scale one evolves the gauge and Yukawa coupling constants,
and the
soft SUSY breaking parameters using renormalization group equations.  As
is well
known, an interesting aspect of supergravity unification is that the
electroweak
symmetry can be broken via renormalization group effects$^{7)}$.  It is
in this
framework that we shall discuss the computation of SUSY particle
spectrum.  The
breaking of the electroweak symmetry is controlled by an effective
potential which
has the form $V = V_0 + \Delta V_1$ where $V_0$ is the renormalization
group
improved tree potential and $\Delta V_1$ is the one-loop effective
potential$^{(8)}$.  Assuming charge and colour conservation, $V_0$ is
given by
 \bea
V_0 &=& m^2_1(t)|H_1|^2 + m^2_2(t) |H_2|^2 - m^2_3(t) (H_1H_2 + {\rm
H.C.}) \nonumber
\\
&& + \frac{1}{8}(g^2_2 + g^2_Y) \left(|H_1|^2 - |H_2|^2 \right)^2
\label{34}
\eea
where $t = \ln(M^2_G/Q^2)$ and $Q$ is the running scale, and $V_1$ is
given by
\beq
V_1 = \frac{1}{64\pi^2} \sum_a (-1)^{2s_a} n_aM^4_a\log
\frac{M^2_a}{e^{3/2}Q^2}
\label{35}
\eeq
The importance of including $V_1$ in the analysis has been emphasized
recently$^{9)}$.  The parameters $m^2_i(t), g_2, g_Y$, satisfy the
following
boundary conditions:
$$
m^2_i(0) = m^2_0 + \mu^2_0~;~~i = 1,2
\eqno{(3.6a)}
$$
$$
m^2_3(0) = -B_0\mu_0
\eqno{(3.6b)}
$$
$$
\alpha_2(0) = (5/3) \alpha_Y(0) = \alpha_G
\eqno{(3.6c)}
$$
Electroweak symmetry breaking requires satisfaction of a number of
conditions.
These include boundedness of the potential from below, i.e., $m^2_1 +
m^2_2 -
2|m^2_3| > 0$ and negativeness of the Higgs (mass)$^2$ matrix, i.e.
$m^2_1m^2_2 -
m^4_3 < 0$.  Minimization of the potential then yields the relations
\addtocounter{equation}{1}
\beq
\frac{1}{2} M^2_Z = \frac{\mu^2_1 - \mu^2_2 \tan^2\beta}{\tan^2\beta -
1}~;~~
\sin^2\beta = \frac{2m^2_3}{\mu^2_1 + \mu^2_2}
\label{37}
\eeq
where $\mu^2_i = m^2_i + \Sigma_i, (i = 1,2)$ and $\Sigma_i$ is the loop
correction
arising from $\Delta V_1$: it has the form
\beq
\Sigma^a = \frac{1}{32\pi^2} \sum_i(-1)^{2s_i}n_i~M^2_i \ln \left[
M^2_i/eQ^2
\right] \frac{\partial M^2_i}{\partial v_a^2}
\label{38}
\eeq
where $v_a = \langle H_a \rangle$.  The particles that make the largest
contributions are the stops and the charginos.  For the stops one has
$\Sigma^a_{\tilde t _i}~(i = 1,2)$ where$^{10)}$
\bea
\Sigma^1_{\tilde t_i} & = &
\frac{3 \alpha_2}{8\pi \cos^2\theta_W}~M^2_{\tilde t_i}
\left[ \frac{1}{4} \mp \left\{ \frac{1}{2} \left( m^2_{\tilde t_L} -
m^2_{\tilde t_R}
\right) \left( \frac{1}{2} - \frac{4}{3} \sin^2 \theta_W \right) \right.
\right.
\nonumber \\
&&\left.\left. + \frac{m^2_t \mu}{M^2_Z \sin^2 \beta} (A_tm_0 \tan \beta
\mu)\right\}~\frac{1}{M^2_{\tilde t_2}  - M^2_{\tilde t_1}} \right] \ln
\left(
\frac{M^2_{\tilde t_i}}{eQ^2} \right)
\label{39}
\eea
\bea
\Sigma^2_{\tilde t_i} &=& \frac{3 \alpha_2}{8\pi \cos^2 \theta_W}~M^2_{
\tilde t_i}
\left[ \left( \frac{m^2_t}{M^2_Z \sin^2\beta} - \frac{1}{4}\right)
\mp \left\{ \frac{1}{2} \left( m^2_{\tilde t_L} - m^2_{\tilde t_R}
\right)
\left( - \frac{1}{2} + \frac{4}{3} \sin^2\theta_W \right)\right.
\right.\nonumber
\\ &&\left. \left.  + \frac{m^2_tA_tm_0}{M^2_Z\sin^2\beta} (A_tm_0 +
\mu~ctn \beta)
\right\}  \times \frac{1}{M^2_{\tilde t_2} - M^2_{\tilde t_1}} \right]
\ln
\left( \frac{M^2_{\tilde t_i}}{eQ^2} \right)
\label{310}
\eea
where $m_{\tilde t_i}$ are the stop masses and $\tilde t_{L,R}$ are
defined in
Section~4.  [Equations (3.9) and (3.10) include a colour factor of three
in
squark contributions missing in Ref.~10].  The chargino contributions
are given
by$^{10)}$
\beq \Sigma^a_{\tilde W_i} = -\frac{\alpha_2 s_i}{2\pi}~
\frac{\lambda^3_i (\lambda_i - u^a\lambda_j)}{(\lambda^2_i -
\lambda^2_j)} \ln
\left( \frac{M^2_{\tilde W_i}}{eQ^2} \right) \quad i,j = 1,2~; \quad i
\not= j
\label{311}
\eeq
where $\lambda_i$ are the eigenvalues of the chargino mass matrix (see
Section~4)
and $u^a = (\tan\beta, \cot \beta)$.

In the analysis of the SUSY spectrum one imposes several constraints
both
theoretical and experimental.  We list the full list of constraints
below:
\begin{itemize}
\item[i)] charge and colour conservation at the electroweak scale and at
the GUT
scale;
\item[ii)]
absence of tachyonic particles;
\item[iii)] a lower bound on SUSY particle masses as indicated by CDF,
DO and
LEP data;
\item[iv)] an upper limit on SUSY masses from naturalness criterion
which we assume
as follows:  $m_0, m_{\tilde g} < 1~{\rm TeV}$;
\item[v)] proton lifetime satisfies the current experimenta bounds;
\item[vi)] the neutralino relic density satisfies a constraint
consistent with the
COBE data.
\end{itemize}
We shall discuss analyses of the SUSY spectrum both including and
excluding
constraints v) and vi) listed above.  Before proceeding further we
discuss the
constraints v) and vi) in some detail.

\noindent
{\it v) Proton stability constraint}

Proton decay is a generic feature of a class of GUT models and string
models, and
thus the current experimental limits on proton lifetime act as a
constraint on the
model.  In analyses discussed below we shall assume that the SUSY
theories we
consider have an SU(5) embedding and the Higgs doublets are embedded in
$5+\bar 5$ representations of SU(5).  In this case, there is a model
independent proton decay amplitude that arises from the exchange of the
Higgs
triplet fields.  The dominant decvay mode of the proton from this
amplitude is the
mode$^{11)}$ $p \rightarrow \bar\nu K^+$ and the total decay width for
this mode is
given by
\beq
\Gamma (p \rightarrow \bar \nu K^+) = \sum_{i=e,\mu,\tau} \Gamma(p
\rightarrow
\bar \nu_i K^+)
\label{312}
\eeq
The contribution of the first generation is essentially negligible.  For
the
remaining two generations one gets the following relation$^{12)}$
\beq
\Gamma(p \rightarrow \bar\nu_i K^+) = C \left( \frac{\beta_p}{M_{H_3}}
\right)^2
|A|^2 |B_i|^2
\label{313}
\eeq
where $A$ depends on quark masses and mixings and is given by
\beq
A = \frac{\alpha^2_2}{2M^2_W} m_sm_c V^\dagger_{21}V_{21}A_LA_S
\label{314}
\eeq
where $V_{ij}$ are the CKM matrix elements and $A_{L,S}$ are the
suppression
factors with values $A_L = 0.283, A_S = 0.833$.  $C$ is a chiral
Lagrangian factor
given by
\beq
C = \frac{m_N}{32\pi f^2_\pi} \left| \left( 1 + \frac{m_N(D+F)}{m_B}
\right) \left(
1 - \frac{m^2_K}{m^2_N} \right) \right|^2
\label{315}
\eeq
where $f_\pi = 139~{\rm MeV}, D = 0.76, F = 0.48$ and $m_B = 1154~{\rm
MeV}$.
$\beta_p$ is the three-quark matrix element of the proton defined by
\beq
\epsilon_{abc} \epsilon_{\alpha\beta}
\langle 0|d^\alpha_{aL} u^\beta_{bL} u^\gamma_{cL}| p \rangle =
\beta_pU^\gamma_L
\label{316}
\eeq
where $U^\gamma_L$ is the proton-wave function.  Recent lattice gauge
calculations
give the following evaluation for $\beta_p~^{13)}$:
\beq
\beta_p = (5.6 \pm 0.8) \times 10^{-3}~{\rm GeV}^3
\label{317}
\eeq
Finally $B_i$ in Eq.~(\ref{313}) is a dressing loop given by$^{12)}$
\beq
B_i = \frac{m^d_i V^\dagger_{i1}}{m_s V^\dagger_{21}}
\left[ P_2B_{2i} + \frac{m_tV_{31}V_{32}}{m_c V_{21}V_{22}} P_3B_{3i}
\right]
\frac{1}{\sin 2\beta}
\label{318}
\eeq
where $P_2, P_3$ are the intergenerational phases which on CP conserving
manifolds have values, $\pm1$ and $B_{ij}$ is the contribution of the
$j^{\rm th}$
generation to $B_i$ and can be written as $B_{ji} = F(\tilde u_i, \tilde
d_j,
\tilde W) + (\tilde d_j \rightarrow \tilde e_j)$.

The 2nd generation contribution $B_{2i}$ where one neglects L-R mixing
is quite
straightforward.  It is given by $B_{2i} = F_2(\tilde c, \tilde d_i,
\tilde W) +
(\tilde d_i \rightarrow \tilde \ell_i)$ where
\beq
F_2(\tilde c, \tilde d_i, \tilde W) =
\sin \gamma_+ \cos \gamma_- f(\tilde c, \tilde d_i, \tilde W_1) +
\cos \gamma_+ \sin \gamma_- f(\tilde c, \tilde d_i, \tilde W_2)
\label{319}
\eeq
and where
\beq
f(a,b,c) = \frac{m_c}{(m^2_b-m^2_c)} \left[ \left[
\frac{m^2_b}{(m^2_a-m^2_b)} \right] \ln \left( \frac{m^2_a}{m^2_b}
\right) - (b
\rightarrow c) \right]
\label{320}
\eeq
In Eq.~(\ref{319}) $\gamma_\mp = \beta _+ \mp \beta_-$ and $\beta_\pm$
are defined by
\beq
\sin 2 \beta_\pm = \frac{(\mu \mp \tilde m_2)}{[4\nu^2_\pm + (\mu \pm
\tilde
m_2)^2]^{1/2}}~;\quad \sqrt 2~\nu_\pm = M_W(\sin \beta \pm \cos \beta)~.
\label{321}
\eeq
The contributions of the third generation involve L-R mixing due to the
top quark
mass and are more complex.  They are discussed in detail in Ref.~12).
For the
purpose of analysing the $p$-stability constraint it is useful to
introduce the
quantity $B$ defined by$^{14,15}$
\beq
B \equiv \left[ |B_2|^2 + |B_3|^2 \right]^{1/2}
\left[ \frac{M_S}{10^{2.4}~{\rm GeV}} \right]^{0.33} \times 10^6~{\rm
GeV}^{-1}
\label{322}
\eeq
where $M_S$ is the effective SUSY mass that appears in Amaldi et al.
type analyses
in fitting the LEP data to SU(5) type SUSY GUT. The multiplicative
factor with
the $M_S$ term takes into account the anticorrelation between $M_S$ and
$M_G$ in
this fit.  The current experimental lower limit on $\bar \nu K^+$ mode
from
Kamiokande is$^{16)}$
\beq
\tau(p \rightarrow \bar \nu K^+) > 1.0 \times 10^{32}~yr
\label{323}
\eeq
Using the above limit one finds$^{14,15)}$
\beq
B \leq 100 \left( \frac{M_{H_3}}{M_G} \right)~{\rm GeV}^{-1}
\label{324}
\eeq
A reasonable upper bound on $M_{H_3}$ is $M_{H_3} \leq 10~M_G$.  Thus
upper limit
keeps the GUT Yukawa couplings perturbative and also keeps $M_{H_3}$
significantly below the Planck scale so that quantum gravity effects
will be
negligible.

\noindent
{\it vi) Neutralino Relic Density Constraint}

In SUSY theories with $R$-parity conservation, the lightest
supersymmetric particle
(LSP) is stable and would contribute to the matter density of the
Universe.  In
supergravity unified theories with radiative breaking one
finds$^{17,18)}$ that for
a large part of the parameter space LSP in fact is the lightest
neutralino $(\tilde
Z_1)$.  Thus in such situations $\tilde Z_1$ is a natural candidate for
cold dark
matter.  If one assumes the inflationary scenario with $\Omega = 1$
(where $\Omega
= \rho/\rho_c$ with $\rho$ the matter density of the Universe and
$\rho_c$ being
the critical matter density needed to close the Universe).  The COBE
data
is consistent with a mix of cold and hot dark matter in the ratio of
2:1. Then one
finds
\beq
0.1 < \Omega h^2 < 0.35
\label{325}
\eeq
where $h$ is the Hubble constant in units of 100km/s.Mpc and lies in the
range
$0.5 < h < 0.75$.  The imposition of the constraint of Eq.~(\ref{325})
requires
considerable care.  The reason for this is that the theoretical
evaluation of
$\Omega h^2$ is a very delicate affair when one is close to thresholds
and poles in
the annihilation cross-section.  We discuss this issue more concretely
below.

The standard formula for the computation of the relic density is given
by$^{19,20)}$
\beq
\Omega_{\tilde Z_1}h^2 \simeq 2.53 \times 10^{-11}
\left( \frac{T_{\tilde Z_1}}{T_\gamma} \right)^3
\left(\frac{T_\gamma}{2.75} \right)^3 ~ \frac{N_f^{1/2}}{J(x_f)}
\label{326}
\eeq
where $(T_{\tilde z_1}/T_\gamma)^3$ is a reheating factor,  $T_\gamma$
is the
current microwave temperature, $N_f$ is the number of degrees of freedom
and
$J(x_f)$ is given by
\beq
J(x_f) = \int^{x_f}_0 dx \langle \sigma v \rangle
\label{327}
\eeq
where $\sigma$ is the annihilation cross-section of neutralinos and $v$
is their
relative velocity.  In usual analyses one uses an expansion on $\sigma
v$ of the
form$^{21-23)}$
\beq
\sigma v = a + bv^2
\label{328}
\eeq
However, it is known$^{24)}$ that Eq.~(\ref{328}) is a poor
approximation in the
neighbourhood of threshold and poles.  Precisely this situation arises
for the case
of annihilation of neutralinos in supergravity models with masses
computed via
radiative breaking.  Here one finds that in the physically interesting
domain of
the parameter space annihilation of neutralinos occurs near Higgs and
$Z$-poles$^{17,18)}$.  In this circumstance the expansion of
Eq.~(\ref{328}) no
longer holds.  However, it turns out that it is fairly straightforward
to carry out
the correct thermal averaging in the presence of poles.  A technique for
accomplishing this is discussed in Refs.~17) and 18).  A similar
analysis is also
discussed in Ref.~25).

\section{SUSY Particle Masses in Supergravity Unification}
\setcounter{equation}{0}
In renormalization group analyses of the SUSY particle spectrum one
begins by
extracting the GUT parameters $\alpha_G, M_G$ by using the two-loop
renormalization
group equations of the gauge coupling constants and fitting to the high
precision
LEP resulfs for $\alpha_i(M_Z), i = 1,2,3$.  The two-loop evolution
equations
are$^{26}$
\beq
\frac{d}{dt} \alpha_i = -\frac{1}{4\pi} \left[ b_i + \frac{1}{4\pi}
\sum_j b_{ij}{\alpha_j} \right] \alpha^2_i
\label{41}
\eeq
where
\bea
b_i &=& (0,-6,-9) + (2,2,2)N_F + (\frac{3}{10},\frac{1}{2},0)N_H
\nonumber \\
\\
b_{ij} &=&
\pmatrix{
0 & 0 & 0 \cr
0 & -24 & 0 \cr
0 & 0 & -54 }
+ \pmatrix{
\frac{38}{15} & \frac{6}{5} & \frac{88}{15} \cr
\frac{2}{5} & 14 & 8 \cr
\frac{11}{15} & 3 &\frac{68}{3} } N_F +
\pmatrix{
\frac{9}{50} & \frac{9}{10} & 0 \cr
\frac{3}{10} & \frac{7}{2} & 0 \cr
0 & 0 & 0 } N_H
\label{43}
\eea
The $\alpha_i$ of Eq.~(\ref{41}) satisfy the boundary conditions
$\alpha_i =
\alpha_G$ at scale $M_G$.  The $\alpha_i$ computed from Eq.~(\ref{41})
are fitted to
the LEP data for $\alpha_i$.  The current determinations for these
are$^{27-29)}$
\beq
\alpha_1(M_Z) \equiv \left( \frac{5}{3} \right) \alpha_Y(M_Z) = 0.016985
\pm 0.00002
\label{44}
\eeq
\beq
\alpha_2(M_Z) = 0.03358 \pm 0.0001
\label{45}
\eeq
\beq
\alpha_3(M_Z) = 0.118 \pm 0.007
\label{46}
\eeq
A fit to the above data$^{30)}$ gives the following values for
$\alpha_G, M_G$ and
$M_S$:
$$
\alpha^{-1}_G = 25.4 \pm 1.7~, \quad M_G \cong
10^{16.2+5.7(\alpha_3/0.118-1)}
\eqno{(4.7a)}
$$
$$
M_S \cong 10^{2.4 + 17.4(1 -\alpha_3/0.118)}
\eqno{(4.7b)}
$$
The analysis of the soft SUSY breaking parameters and of the
$\mu$-parameter is
done using one-loop renormalization group equations$^{31)}$.  The
gaugino masses
are assumed to obey the $RG$ equation
\addtocounter{equation}{1}
\beq
\frac{d \tilde m_i}{dt} = - \frac{b_i}{4\pi} \tilde \alpha_i(t) \tilde
m_i(t)~;
\quad
\tilde m_i(0) = m_{1/2}
\label{48}
\eeq
The $\mu$-parameter and the top Yukawa coupling obey the equation$^{7)}$
\beq
\frac {d\mu^2}{dt} = \left( 3 \tilde \alpha_2 + \frac{3}{5} \tilde
\alpha_1 - 3Y_t
\right) \mu^2
\label{49}
\eeq
\beq
\frac{dY_t}{dt} = \left( \frac{16}{3} \tilde \alpha_3 + 3 \tilde
\alpha_2 +
\frac{13}{15} \tilde \alpha_1 - 6 Y_t \right) Y^2_t
\label{410}
\eeq
where $Y_t = h^2_t/4\pi$ and $h_t$ is the top quark Yukawa coupling.
The chargino
masses are determined completely in terms of $\mu$ and $\tilde
m_2~^{1)}$:
\beq
\lambda_i = \frac{1}{2} \left( |4 \nu^2_+ + (\mu - \tilde m _2)^2|^{1/2}
\mp
|4 \nu^2_- + (\mu + \tilde m_2)^2|^{1/2} \right)
\label{411}
\eeq
where $\sqrt 2 \nu_{\pm} = M_W (\sin \beta \pm \cos \beta)$ and
$m_{\tilde W_i} =
|\lambda_i|, i = 1,2$.  Similarly the neutralino masses are given by
roots of the
secular equation $f(\lambda) = 0$ where$^{1),31)}$
\bea
f(\lambda) & \equiv & \lambda^4 - (m_{\tilde \gamma} + m_{\tilde Z})
\lambda^3 -
(M^2_Z + \mu^2 + m^2_{\tilde \gamma \tilde Z} - m_{\tilde \gamma}
m_{\tilde Z})
\lambda^2 \nonumber \\
& + & \left[ ( m_{\tilde \gamma} - \mu \sin 2\beta)M^2_Z + (m_{\tilde
\gamma} +
m_{\tilde Z})\mu^2 \right] \lambda  \nonumber \\
&+& (\mu^2(m^2_{\tilde \gamma \tilde Z} - m_{\tilde \gamma}m_{\tilde Z})
+ \mu
m_{\tilde \gamma} M^2_Z \sin 2 \beta)
\label{412}
\eea
where $m_{\tilde \gamma}, m_{\tilde Z}$ and $m_{\tilde \gamma \tilde z}$
are as
defined in the first paper of Ref.~1.

 {\it Sleptons~(i = 1,2,3)}
Masses of the three generation of sleptons are given in a
straightforward fashion
and are as follows$^{1)}$:
\beq
m^2_{\tilde e_{iL}} = m^2_0 + m^2_{ei} + \tilde \alpha_G
\left[ \left(\frac{3}{2}\right)f_2 + \left(\frac{3}{10}\right)f_1
\right]M^2_{1/2} +
\left( -\frac{1}{2} + \sin^2\theta_W \right) M^2_Z \cos 2\beta
\label{413}
\eeq
\beq
m^2_{\tilde e iR} = m^2_0 + m^2_{e_i} + \tilde \alpha_G
\left( \frac{6}{5} \right) f_1m^2_{1/2} - \sin^2 \theta_WM^2_Z \cos
2\beta
\label{414}
\eeq
\beq
m^2_{\tilde \nu_{iL}} = m^2_0 + \tilde \alpha_G \left[ \left(
\frac{3}{2} \right)f_2
+ \left( \frac{3}{10} \right)f_1 \right] m^2_{1/2} + \left( \frac{1}{2}
\right)
M^2_Z \cos 2 \beta
\label{415}
\eeq
where $\tilde \alpha_G = \alpha_G/4\pi, f_a(t) = t(2-\beta_at)/(1 +
\beta_at)^2$ and
where $\beta_a = (33/5, 1, -3)$

{\it Squarks (i = 1,2)}: for the first two generation of squarks one can
ignore the
left-right mixing since the quark masses are small.  In this
approximation masses of
the first two squark generations are given by$^{1)}$
\beq
m^2_{\tilde u_{iL}} = m^2_0 + m^2_{u_i} + \tilde \alpha_G
\left[ \left( \frac{8}{3}\right)f_3 + \left( \frac{3}{2}\right)f_2 +
\left( \frac{1}{30} \right) f_1 \right] m^2_{1/2} +
\left( \frac{1}{2} - \frac{2}{3} \sin^2 \theta_W \right) M^2_Z \cos 2
\beta
\label{416}
\eeq
\beq
m^2_{\tilde d_{iL}} = m^2_0 + m^2_{d_i} + \tilde \alpha_G
\left[ \left( \frac{8}{3}\right)f_3 + \left( \frac{3}{2}\right)f_2 +
\left( \frac{1}{30} \right) f_1 \right] m^2_{1/2} +
\left( -\frac{1}{2} + \frac{1}{3} \sin^2 \theta_W \right) M^2_Z \cos 2
\beta
\label{417}
\eeq
\beq
m^2_{\tilde u_{iR}} = m^2_0 + m^2_{u_i} + \tilde \alpha_G
\left[ \left( \frac{8}{3}\right)f_3 + \left( \frac{8}{15}\right)f_1
\right] m^2_{1/2}
+ \left( \frac{2}{3}\right)\sin^2 \theta_W  M^2_Z \cos 2 \beta
\label{418}
\eeq
\beq
m^2_{\tilde d_{iR}} = m^2_0 + m^2_{d_i} + \tilde \alpha_G
\left[ \left( \frac{8}{3}\right)f_3 + \left( \frac{2}{15}\right)f_1
\right] m^2_{1/2}
+ \left( -\frac{1}{3}\right)\sin^2 \theta_W  M^2_Z \cos 2 \beta
\label{419}
\eeq
An interesting feature of supergravity masses is that at any scale
\beq
m^2_{\tilde c} - m^2_{\tilde u} = m^2_c - m^2_u
\label{420}
\eeq
Eq.(\ref{420}) thus leads to a natural suppression of flavour changing
neutral
currents$^{4)}$.

{\it Squarks (i = 3)}:  the $t$-squark masses are affected significantly
due to the top
mass.  There is a significant amount of L-R mixing and the
stop(mass)$^2$ are given by
eigenvalues of the following (mass)$^2$ matrix$^{1)}$:
\beq
\pmatrix{
m^2_{\tilde t_L} & m_t(A_t + \mu \cot \beta) \cr
m_t(A_t+\mu\cot \beta) & m^2_{\tilde t_R}}
\label{421}
\eeq
where$^{1,7)}$
\beq
m^2_{\tilde t_L} = m^2_Q + m^2_t + \left[ \left( -\frac{1}{2}\right) +
\left(\frac{2}{3}\right) \sin^2 \theta_W \right] M^2_Z \cos 2 \beta
\label{422}
\eeq
\beq
m^2_{\tilde t_R} = m^2_U + m^2_t + \left[ \left( -\frac{2}{3}\right)
\sin^2 \theta_W \right] M^2_Z \cos 2 \beta
\label{423}
\eeq
\beq
m^2_U = \frac{1}{3} m^2_0 + \frac{2}{3} f A_0m_{1/2} - \frac{2}{3}
kA^2_0 +
\frac{2}{5} hm^2_0 + \left[ \frac{2}{3} e + \tilde \alpha_G \left(
\frac{8}{3} f_3 - f_2 + \frac{1}{3} f_1 \right) \right] m^2_{1/2}
\label{424}
\eeq
\beq
m^2_Q = \frac{2}{3} m^2_0 + \frac{1}{3} f A_0m_{1/2} - \frac{1}{3}
kA^2_0 +
\frac{1}{3} hm^2_0 + \left[ \frac{1}{3} e + \tilde \alpha_G \left(
\frac{8}{3} f_3 + f_2 - \frac{1}{15} f_1 \right) \right] m^2_{1/2}
\label{425}
\eeq
where the functions $e, f, h, k$ are as defined in Ib\'a$\tilde{\rm
n}$ez et
al.$^{32)}$.  The $\tilde b_R$ mass is given by Eq.~(\ref{419}) as for
the two
generation case, while the $\tilde b_L$ mass is modified by third
generation effects
as is given by
\beq
m^2_{\tilde b L} = \frac{1}{2} m^2_0 + m^2_b + \frac{1}{2} m^2_{\tilde
U} + \tilde
\alpha_G \left[ \left(\frac{4}{3} \right)f_3 + \left( \frac{1}{15}
\right)f_1
\right] m^2_{1/2} + \left( - \frac{1}{2} + \frac{1}{3} \sin^2 \theta_W
\right) m^2_Z
\cos 2 \beta
\label{426}
\eeq

\noindent
{\it Higgs}:  The parameters of the Higgs potential are given by the
following
evolution equations at the one-loop level
\beq
m^2_1(t) = m^2_0 + \mu^2(t) + gm^2_{1/2}
\label{427}
\eeq
\beq
m^2_t(t) = \mu^2(t) + e(t)m^2_{1/2} + A_0m_{1/2}f + m^2_0h - A^2_0k
\label{428}
\eeq
\beq
m^2_3 = -B_0\mu^2(t) + r\mu_0m_{1/2} + sA_0\mu_0
\label{429}
\eeq
where $g, r, s$ are as defined in Ref.~32).  The Higgs masses using just
the tree
potential are given by$^{1)}$
\beq
m^2_H = m^2_A + M^2_W
\label{430}
\eeq
\beq
m^2_A = m^2_1 + m^2_2 = \frac{2m^2_3}{\sin 2\beta}
\label{431}
\eeq
\beq
m^2_{h,H} = \frac{1}{2} \left[ M^2_Z + M^2_A \mp \left\{ (M^2_Z +
m^2_A)^2 -
4m^2_AM^2_Z \cos 2\beta \right\} ^{1/2} \right]
\label{432}
\eeq
However, there can be significant corrections from the loop effects.  We
discuss
here the corrections to the neutral Higgs sector.  Retaining only the
top Yukawa
coupling one has$^{33,34)}$
\bea
m^2_A &=& \frac{\Delta}{\sin^2\beta} \nonumber \\
\Delta &=& 2m^2_3 - \frac{3\alpha_2}{8 \pi} \frac{\mu A_t}{\sin^2\beta}
\left( \frac{m_t}{M_W} \right)^2 \left( \frac{f(m^2_{\tilde t_1}) -
f(m^2_{\tilde
t_2})} {m^2_{\tilde t_1} - m^2_{\tilde t_2}} \right)
\label{433}
\eea
where $f(m^2) = 2m^2(\log(m^2/Q^2)-1)$.  The values of $m_{h,H}$ are
also modified.
One has$^{33)}$
\beq
m^2_{h,H} = \frac{1}{2} \left[ M^2_Z + m^2_A + \epsilon \mp
\left\{ (M^2_Z + m^2_A + \epsilon)^2 - 4m^2_AM^2_Z \cos 2\beta +
\epsilon_1
\right\}^{1/2} \right]
\label{434}
\eeq
where $\epsilon, \epsilon_1$ are the loop corrections which are given by
\beq
\epsilon = Tr~\Delta~; ~~ \epsilon_1 = -4(Tr~\nu \Delta + \det \Delta)
\label{435}
\eeq
and $\nu$ and $\Delta$ are defined by
\bea
\nu_{11} &=& s^2M^2_Z + c^2m^2_A~;~~\nu_{22} = c^2M^2_Z +
s^2m^2_A~;~~\nu_{12} =
\nu_{21} = sc(M^2_Z + m^2_A) \nonumber \\
\Delta_{11} &=& x\mu^2y^2z~;~~\Delta_{12} = x\mu y(w + A_t yz) =
\Delta_{21}
\nonumber \\
\Delta_{22} &=& x(v + 2A_t yw + A^2_t y^2 z)
\label{436}
\eea
where
\bea
x &=& \frac{3\alpha^2}{4\pi} \frac{m^4_t}{M^2_Ws^2}~;~~
y = \frac{A_t + \mu ctn \beta}{m^2_{\tilde t_1}-m^2_{\tilde t_2}}~;~~
z=2 - w \frac{m^2_{\tilde t_1} + m^2_{\tilde t_2}}{m^2_{\tilde t_1} -
m^2_{\tilde
t_2}} \nonumber \\
w &=& \ln \left( \frac{m^2_{\tilde t_1}}{m^2_{\tilde t_2}} \right)~:~~ v
= \ln \left(
\frac{m^2_{\tilde t_1}m^2_{\tilde t_2}}{m^4_t}\right)
\label{437}
\eea
where $(s,c) = (\sin \beta, \cos \beta)$.

\section{Analysis and Results}
\setcounter{equation}{0}
We begin by discussing the parameter space of the supergravity unified
models.  From
Eq.~(\ref{33}) we find that the low-energy theory is defined by the
seven parameters
given there.  As discussed in Section~4, $\alpha_G$ and $M_G$ are
determined using
the LEP data.  Further using the first of the two equations in
Eq.~(\ref{37}).  Thus
the 7-dimensional parameter space reduces to a 4-dimensional parameter
space defined
by
\beq
m_0~,~~m_{1/2}~,~~A_t~,~~\tan \beta
\label{51}
\eeq
where $A_t$ is the value of the trilinear coupling $A_0$ at the
electroweak scale.

One of the interesting features that emerges from the analysis is that
the parameter
$\mu$, which as discussed above is determined by the $E-W$ symmetry
breaking
equation, is typically large$^{14,15,36)}$ in the sense $|\mu| \gg M_Z$
over much of
the parameter space.  The largeness of $\mu$ leads to certain scaling
properties.
For the charginos one finds for $|\mu| \gg M_Z$ the result
\beq
m_{\tilde W_1} \simeq \tilde m_2 -
\frac{M^2_W\sin 2\beta}{\mu}~;~~ m_{\tilde W_2} \cong \mu +
\frac{M^2_W}{\mu}
\label{52}
\eeq
Similarly for neutralinos one has
$$
m_{\tilde Z_1} \cong \tilde m_1 - \frac{M^2_Z \sin 2\beta
\sin^2\theta_W}{\mu}~;~~
m_{\tilde Z_2} \cong \tilde m_2 - \frac{M^2_W\sin 2 \beta}{\mu}
\eqno{(5.3a)}
$$
$$
m_{\tilde Z_3, \tilde Z_4} \cong
\left| \mu -\frac{1}{2} \frac{M^2_Z}{\mu} (1 \pm \sin^2\beta) \right|
\eqno{(5.3b)}
$$

{}From Eqs.~(5.2) and (5.3) one finds the following scaling
relations$^{14,15,35)}$
$$
m_{\tilde W_1} \simeq 2m_{\tilde Z_1} \simeq m_{\tilde Z_2}
\eqno{(5.4a)}
$$
$$
m_{\tilde W_2} \simeq m_{\tilde Z_3} \simeq m_{\tilde Z_4}
\eqno{(5.4b)}
$$
The numerical analysis exhibits the above scaling relations.  Also to a
good
approximation the following relations between the chargino and gluino
masses
emerge:
 \addtocounter{equation}{2}
\beq
m_{\tilde W_1} \cong \frac{1}{4} m_{\tilde g}(\mu > 0)~;~~m_{\tilde W_1}
\simeq
\frac{1}{3} m_{\tilde g} \quad(\mu < 0)
\label{55}
\eeq

There is a scaling relation in the Higgs sector also.  Here one finds
\beq
m_{H^0} \simeq m_A \simeq m_{H^\pm}
\label{56}
\eeq
Next we discuss the result of the analysis for four specific models.
These are (1)
models with constraints (i)--(iv), (2) models with constraints (i)--(v),
(3)
models of type (1) with constraint (vi) and (4) models of type (2) with
constraints (vi).

(1) These models (sometimes referred to as MSSM) are
a truncated version of the supergravity unification where proton decay
via
dimension five operators is discarded or suppressed.  Thus here only the
constraints
(i)--(iv) are imposed$^{36)}$.  Due to the absence of constraints (v)
and (vi)
$m_{\tilde g}$ here can run up to its naturalness limit.  The
characteristic
spectrum after one integrates over the full parameter space of the
theory is shown
in Fig.~1.  One finds  a wide dispersion in the spectrum from a lower
limit of
0(25~GeV) for the lightest neutralino to an upper limit of well above a
TeV for
charged Higgs.  An interesting feature of the spectrum is that the
lightest
neutral Higgs has an upper limit on its mass of about 130~GeV.  However,
the
lightest chargino in this scenario can be as large as 300~GeV.  The
lower limits
on masses of $\tilde e_R, \tilde \nu_L, \tilde t_1, \tilde Z_1, \tilde
Z_2, \tilde
W_1$ and $h^0$ are set only by experiment.  The lower limits for the
remaining
particles lie typically in the region 100--200~GeV.  Specifically one
finds that the
mass of the charged Higgs is greater than 100~GeV.

(2)
Here one imposes constraints (i)--(v) and the model corresponding to the
minimal
supergravity grand unification.  The characteristic spectrum assuming
$M_{H_3} \leq
10M_G$ (and hence from Eq.~(\ref{324}), $B \leq 1000~{\rm GeV}^{-1}$) is
given in
Fig.~2 when one integrates over the full parameter space of the model.
An
interesting feature of the model is that because of proton stability
requirement
$m_0 > m_{\tilde g}$, and since $m_{\tilde g}$ has an experimental lower
bound of
$\approx~150~{\rm Gev}$, it implies that the heavier Higgs $H^0, A,
H^\pm$ cannot
be too light.  In fact the detailed analysis shows that the $H^0, A,
H^\pm$ masses
are always larger than 200~Gev.  In this context we may recall that
there is a
so-called ``hole" in the Higgs mass range of 100--200~GeV $(5 \leq \tan
\beta \leq
20)$ which cannot be probed at LEP and LHC$^{37)}$.  In the context of
supergravity
grand unification discused here, such a ``hole" is excluded.  The
light Higgs mass obeys the limit $m_{h^0} \leq 130~{\rm GeV}$ (see
Figs.~3 and 4).  As
for the case of $H^0, A$ and $H^\pm$ masses, the lower limits of the
sleptons and
squark masses (except for the stop1 mass) and also the lower limits of
$\tilde Z_3,
\tilde Z_4$ and $\tilde W_2$ masses are pushed up above 200~GeV.  These
results can be
easily understood from the proton stability constraint which requires
$m_0 > m_{\tilde
g}$, and leads to larger lower limits for the slepton and squark masses.
Also $m_0 >
m_{\tilde g}$ implies a larger lower limit on $|\mu|$ in radiative
electroweak
breaking and thus leads to larger lower limits on $\tilde Z_3, \tilde
Z_4$ and $\tilde
W_2$ masses.

(3) Supergravity Unification with Dark Matter Constraints:  here we
impose the
constraints (i)--(iv) and (vi).
In this case one finds an upper limit on the gluino mass of $m_{\tilde
g} \lappeq
800~{\rm GeV}$.  Also one finds an upper limit on the lighter chargino
mass,
$M_{\tilde W_1} \lappeq 250~{\rm GeV}$.  The mass spectrum with
integration over
the full parameter space under the constraints (i)--(iv) and (vi) is
exhibited in
Fig.~5.  The lower limits on masses of the three generation sleptons and
squarks are
typically similar, though somewhat larger, as for case (1).

(4) Supergravity Unification with Proton Stability and Relic Density:
for this
case the full set of constraints (i)--(vi) are imposed.  Here we find an
upper
limit on the gluino mass of $m_{\tilde g} \leq$ 400~GeV, and an upper
limit on the
chargino mass of $M_{\tilde W_1} \leq$ 120~GeV.  The mass spectrum after
integration over the full parameter space under the constraints
(ii)--(vi) is
shown in Fig.~6.  Here the lower limits on masses of the three
generation of
sleptons and squarks (except stop 1) and on masses of $\tilde Z_3,
\tilde Z_4,
\tilde W_2$, and of $H^0, A, H^\pm$ Higgs are substantially higher than
those for
cases (1) and (3), and are similar to those for case (2).  An
interesting result of
the analysis is that the mass of the lightest chargino is typically
smaller than
the mass of the lightest higgs, i.e., $m_{\tilde W_1} \leq m_{h^0}$.

\section{Conclusion}

SUSY spectrum is discussed within the framework of supergravity unified
theories.
There are 32 SUSY particles in these theories whose masses can be
predicted in terms
of 4 parameters.  Thus there are 28 predictions some of which can be
translated
also in terms of sum rules$^{38)}$.  It is found that the SUSY spectrum
exhibits
certain scaling properties over much of the parameter space of the
theory.
Computation of the SUSY spectrum is carried out for four different
scenarios:
supergravity grand unification without proton decay, supergravity grand
unification
including proton stability, supergravity unification with neutralino
relic
density constraint but without proton stability constraint, and
supergravity
unification with neutralino relic density and proton stability
constraints.  The
analysis of supergravity grand unification with proton stability
constraint shows
that the so-called ``hole" in the CP odd Higgs mass between 100--200~GeV
which cannot
be explored experimentally at LEP2 and LHC, is eliminated.

One also finds some interesting features in the spectrum when the dark
matter
constraint is included in the analysis.  One of these is the observation
that
$m_{\tilde g} \lappeq 800~{\rm GeV}$, and $m_{\tilde W_1} \lappeq
250~{\rm GeV}$
over the allowed region of the parameter space.  With inclusion of both
$p$-stability
and dark matter constraints one finds the remarkable result that
$m_{\tilde W_1}
\lappeq 120$~GeV and $m_{\tilde W_1} \lappeq m_{h^0}$.  Also for all the
four scenarios
we find $m_{h^0} \lappeq 130~{\rm GeV}$.  It should be of interest to
pursue signals of
supersymmetry in this context using, for example, the trileptonic signal
in off-shell
$W$ decays$^{39)}$.

\noindent
{\bf Acknowledgements}

This research was supported in part by NSF grant Nos. PHY-19306906 and
PHY-916593.
\vfill\eject

\noindent
{\bf References}
\begin{itemize}
\item[1)]
For a review see:\\
P. Nath, R. Arnowitt and A.H. Chamseddine. ``Applied $N=1$
Supergravity", (World
Scientific, Singapore, 1984);\\
H.P. Nilles, {\it Phys. Rep.} {\bf 110} (1984) 1;\\
H. Haber and G. Kane, {\it Phys. Rep.} {\bf C117} (1985) 75.
\item[2)] A.H. Chamseddine, R. Arnowitt and P. Nath, \PRL {\bf 29}
(1982) 970.
\item[3)] R. Barbieri, S. Ferrara and C.A. Savoy, \PL {\bf B119} (1983)
343.
\item[4)] S. Dimopoulos and H. Georgi, \NP {\bf B193} (1981) 150;\\
N. Sakai, {\it Zeit. f. Phys.} {\bf C11} (1981) 153;
L. Girardello and
M.T. Grisaru, \NP {\bf B194} (1982) 65.
 \item[5)] L. Hall, J. Lykken and S. Weinberg,
\PR {\bf D22} (1983) 2359. \item[6)] P. Nath, R. Arnowitt and A.H.
Chamseddine, \NP
{\bf B227} (1983) 121;\\ S. Soni and A. Weldon, \PL {\bf B216} (1983)
215.
\item[7)] K. Inoue et al., {\it Progr. Theor. Phys.} {\bf 68} (1982)
927;\\
L. Ib\'a$\tilde{\rm n}$ez and G.G. Ross, \PL {\bf 110} 227 (1982);\\
L. Alvarez-Gaum\'e, J. Polchinski and M.B. Wise, \NP {\bf B250} (1983)
495;\\
J. Ellis, J. Hagelin, D.V. Nanopoulos and K. Tamvakis, \PL {\bf 125B}
(1983)
225;\\
L.E. Ib\'a$\tilde{\rm n}$ez and C. Lopez, \PL {\bf 128} (1983) 2887.
\item[8)] S. Coleman and E. Weinberg, \PR {\bf D7} (1973) 1888;\\
S. Weinberg, \PR {\bf D7} (1973) 2887.
\item[9)] G. Gamberini, G. Ridolfi and F. Zwirner, \NP {\bf B331} (1990)
331.
\item[10)] R. Arnowitt and P. Nath, \PR {\bf D46} (1992) 725.
\item[11)] S. Weinberg, \PR {\bf D26} (1982) 287;\\
N. Sakai and T. Yanagida, \NP {\bf B197} (1982) 533;\\
S. Dimopoulos, S. Raby and F. Wilczek, \PL {\bf B112} (1982) 133;\\
J. Ellis, D.V. Nanopoulos and S. Rudaz, \NP {\bf B202} (1982) 43;\\
S. Chadha and M. Daniels, \NP {\bf B229} (1983) 105.
\item[12)] R. Arnowitt, A.H. Chamseddine and P. Nath, \PL {\bf B156}
(1985) 215;\\
P. Nath, R. Arnowitt and A.H. Chamseddine, \PR {\bf D32} (1985) 2348;\\
J. Hisano, H. Murayama and T. Yanagida, \NP {\bf B402} (1993) 46.
\item[13)] B. Gavela et al., \NP {\bf B312} (1989) 269.
\item[14)] R. Arnowitt and P. Nath, \PRL {\bf 69} (1992) 368.
\item[15)] P. Nath and R. Arnowitt, \PL {\bf B289} (1992) 368.
\item[16)] R. Becker-Szendy et al., \PR {\bf D47} (1993) 4028.
\item[17)] R. Arnowitt and P. Nath, \PL {\bf B299} (1993) 58 and Erratum
ibid
{\bf B303} (1993) 403.
\item[18)] P. Nath and R.Arnowitt, \PRL {\bf 70} (1993) 3696.
\item[19)] See, e.g., J. Ellis, J.S. Hagelin, D.V. Nanopoulos, K. Olive
and M.
Srednicki, \NP {\bf B238} (1984) 453;\\
J. Ellis, L. Roszkowski and Z. Lalak, \PL {\bf B245} (1990) 545.
\item[20)] For a review see R. Kolb and M. Turner, ``The Early
Universe",
Addison-Wesley Publishing Company, 1990.
\item[21)] B.W. Lee and S. Weinberg, \PRL {\bf 39} (1977) 165;\\
D.A. Dicus, E. Kolb and V. Teplitz, \PRL {\bf 39} (1977) 168;\\
H. Goldberg, {\it Phys. Rev. Lett.} {\bf 50} (1983) 1419.
\item[22)] J. Lopez, D.V. Nanopoulos and K. Yuan, \NP {\bf B370} (1992)
445.
\item[23)] M. Drees and M.M. Nojiri, \PR {\bf D47} (1993) 376.
\item[24)] K. Griest and D. Seckel, \PR {\bf D43} (1991) 3191;\\
P. Gondolo and G. Gelmini, \NP {\bf B360} (1991) 145.
\item[25)] S. Kelley, J.L. Lopez, D.V. Nanopoulos, H. Pois and K. Yuan,
\PR {\bf
D47} (1993) 2461.
\item[26)] M.B. Einhorn and D.R.T. Jones, \NP {\bf B196} (1982) 475.
\item[27)] M. Davier, Proc. Lepton-Photon High Energy Physics
Conference, Geneva,
1991, Eds. S. Hegarty, K. Potter, E. Quercigh (World Scientific,
Singapore, 1991).
\item[28)] S. Fanchiotti, B. Kniel and A. Sirlin, \PR {\bf D48} (1993)
307.
\item[29)] H. Bethke, XXVI Proc. Conference on High Energy Physics,
Dallas 1992,
Ed. J. Sanford, AIP Conf. Proc. No. 272 (1993);\\
G. Altarelli, talk at Europhysics Conference on High Energy Physics,
Marseille,
August 1993.
\item[30)] P. Langacker, Proc. PASCOS 90 Symposium, Eds. P. Nath andS.
Reucroft,
(World Scientific, Singapore, 1990);\\
J. Ellis, S. Kelley and D.V. Nanopoulos, \PL {\bf 249B} (1990) 441;\\
U. Amaldi, W. de Boer and H. F\"urstenua, \PL {\bf 260B} (1991) 447;\\
F. Anselmo, L. Cifarelli, A. Peterman and A. Zichichi, {\it Nuov. Cim}
{\bf 104A}
(1991) 1817; {\bf 115A} (1992) 581;\\
W. de Boer, Karlsruhe preprint IEKP-KA/94-01 (1994).
\item[31)] Recently two-loop renormalization group equations for the
soft SUSY
breaking parameters have also been derived, see:\\
S. Martin and M.T. Vaughn, Northeastern Univ. preprint NUB-3081-93TH
(1993);\\
Y. Yamada, KEK preprint, KEK-TH.371 (1993).
\item[32)] L. Ib\`a$\tilde{\rm n}$ez, C. Lopez and C. Mu$\tilde{\rm
n}$oz, \NP
{\bf B256} (1985) 218.
\item[33)] Y. Ikada, M. Yamaguchi and T. Yanagida, {\it Prog. Theor.
Phys.} {\bf
85} (1991) 1;\\
J. Ellis, G. Ridolfi andE. Zwirner, \PL {\bf B257} (1991) 83;\\
H.E. Haber and R. Hampling, \PRL {\bf 66} (1991) 1815.
\item[34)] Two-loop corrections to the Higgs mass have been investigated
recently
by:\\
R. Hempfling and A.H. Hoang, DESY preprint DESY93-162 (1993).
\item[35)] P. Nath and R. Arnowitt, \PL {\bf B287} (1992) 3282.
\item[36)] Similar analyses but with somewhat different naturalness
constraints
have been carried out recently by:\\
G.G. Ross and R.G. Roberts, \NP {\bf B377} (1992) 971;\\
J.L. Lopez, D.V. Nanopoulos and A. Zichichi, CERN-PPE/94-01 (1994);\\
G.L. Kane, C. Kolda, L. Rozskowski and J.D. Wells, Univ. Michigan
preprint
UM-TH-93-24 (1993);\\
V. Barger, M.S. Berger and P. Ohman, Univ. of Wisconsin preprint
MAD/PH/801
(1993).
\item[37)] Z. Kunszt and F. Zwirner, \NP {\bf B385} (1992) 3;\\
H. Baer, C. Kao and X. Tata, \PL {\bf B303} (1993) 289.
\item[38)] S. Martin and P. Ramond, {\it Phys. Rev.} {\bf D48} (1993)
5365.
\item[39)] P. Nath and R. Arnowitt, {\it Mod. Phys. Lett.} {\bf A2}
(1987) 331;\\
R. Barbieri, F. Caravaglios, M. Frigeni and M. Mangano, \NP {\bf B367}
(1991)
28;\\
H. Baer and X. Tata, \PR {\bf D47} (1993) 2739.

\end{itemize}

\vfill\eject
\noindent
{\bf Figure Captions}
\begin{itemize}
\item[Fig. 1:]  Mass ranges of the SUSY mass spectrum in radiative
electroweak
symmetry breaking for MSSM under the natural constraint $m_0, m_{\tilde
g}
\lappeq 1~{\rm TeV}$, for the case $m_t = 160~{\rm GeV}$ and $\mu~<~0$.
The
particles are labelled top to bottom as follows:$\tilde e_L, \tilde e_R,
\tilde \nu, \tilde u_L, \tilde u _R, \tilde d_L, \tilde d_R, \tilde b_L,
\tilde
t_1, \tilde t_2, $ $\tilde Z_1,  \tilde Z_2,  \tilde Z_3, \tilde Z_4,
\tilde W_1, \tilde
W_2, \tilde g, h^0, H^0, A$ and $H^\pm$.
\item[Fig. 2:] Mass ranges of the SUSY masses in
supergravity GUTs including constraint of $p$-stability.  Particles are
labelled
as in Fig.~1.  The parameters are also as in Fig.~1.
\item[Fig. 3:] Light Higgs mass as a function of the dressing loop
function for
$m_t = 160$, and $\mu < 0$ when all other parameters are integrated out
as in
Fig.~2.
\item[Fig.~4:]  Same as Fig. 3 except $\mu > 0$.
\item[Fig. 5:]  Same as Fig. 1 except including dark matter constraint.
\item[Fig. 6:]  Same as Fig. 5 except also including proton stability
constraint.  The
particles are labelled top to bottm as follows:  $\tilde e_L, \tilde
e_R, \tilde \nu,
\tilde u_L, \tilde u_R, \tilde d_L, \tilde d_R, \tilde b_L, \tilde t_1,
\tilde t_2, \tilde Z_1,
\tilde Z_2, \tilde Z_3/\tilde Z_4, \tilde W_1, \tilde W_2, \tilde g,
h^0, H^0, A$ and
$H^\pm$.
 \end{itemize}

\end{document}